


\font\scrsf=cmssi10                

\font\ninebf=cmbx9




\def\gboxit#1{\hbox{\vrule\vbox{\hrule\kern3pt\vtop
{\hbox{\kern3pt#1\kern3pt}
\kern3pt\hrule}}\vrule}}

\def\ttilde#1{\raise2ex\hbox{${\scriptscriptstyle(}\!
\sim\scriptscriptstyle{)}$}\mkern-16.5mu #1}
\def\dddots#1{\raise1ex\hbox{$^{\ldots}$}\mkern-16.5mu #1}
\def\pp#1#2{\raise1.5ex\hbox{${#2}$}\mkern-17mu #1}
\def\upleftarrow#1{\raise1.5ex\hbox{$\leftarrow$}\mkern-16.5mu #1}
\def\uprightarrow#1{\raise1.5ex\hbox{$\rightarrow$}\mkern-16.5mu #1}
\def\upleftrightarrow#1{\raise1.5ex\hbox{$\leftrightarrow$}\mkern-16.5mu #1}
\def\bx#1#2{\vcenter{\hrule \hbox{\vrule height #2in \kern #1in\vrule}\hrule}}

\def\squiggle#1{\lower1.5ex\hbox{$\sim$}\mkern-14mu #1}

\def\narrower{\advance\leftskip by\parindent \advance\rightskip by\parindent}

\def\mbox#1#2{\vcenter{\hrule width#1in\hbox{\vrule height#2in
   \hskip#1in\vrule height#2in}\hrule width#1in}}
\def\eqsquare #1:#2:{\vcenter{\hrule width#1\hbox{\vrule height#2
   \hskip#1\vrule height#2}\hrule width#1}}
\def\inbox#1#2#3{\vcenter to #2in{\vfil\hbox to #1in{$$\hfil#3\hfil$$}\vfil}}
\def\strutdepth{\dp\strutbox}
\def\marbul{\strut\vadjust{\kern-\strutdepth\specialbul}}
\def\specialbul{\vtop to \strutdepth{
    \baselineskip\strutdepth\vss\llap{$\bullet$\qquad}\null}}
\def\Bcomma{\lower6pt\hbox{$,$}}    
\def\bcomma{\lower3pt\hbox{$,$}}    

\def\sl{\scrsf}

\def\updots{\mathinner{\mskip 1mu\raise 1pt\hbox{.}
    \mskip 2mu\raise 4pt\hbox{.}\mskip 2mu
    \raise 7pt\vbox{\kern 7pt\hbox{.}}\mskip 1mu}}
\def\svec#1{\skew{-2}\vec#1}

\def\square{\kern1pt\vbox{\hrule height 1.2pt\hbox{\vrule width 1.2pt\hskip 3pt
   \vbox{\vskip 6pt}\hskip 3pt\vrule width 0.6pt}\hrule height 0.6pt}\kern1pt}
\def\ssquare{\kern1pt\vbox{\hrule height .6pt\hbox{\vrule width .6pt\hskip 3pt
   \vbox{\vskip 6pt}\hskip 3pt\vrule width 0.6pt}\hrule height 0.6pt}\kern1pt}
\def\lege{\hbox{$ {     \lower.40ex\hbox{$>$}
                   \atop \raise.20ex\hbox{$<$}
                   }     $}  }

\def\rege{\hbox{$ {     \lower.40ex\hbox{$<$}
                   \atop \raise.20ex\hbox{$>$}
                   }     $}  }

\def\lapp{\hbox{$ {     \lower.40ex\hbox{$<$}
                   \atop \raise.20ex\hbox{$\sim$}
                   }     $}  }
\def\rapp{\hbox{$ {     \lower.40ex\hbox{$>$}
                   \atop \raise.20ex\hbox{$\sim$}
                   }     $}  }

\def\tridots{\hbox{$ {     \lower.40ex\hbox{$.$}
                   \atop \raise.20ex\hbox{$.\,.$}
                   }     $}  }
\def\Times{\times\hskip-2.3pt{\raise.25ex\hbox{{$\scriptscriptstyle|$}}}}

\def\rightonleft{\hbox{$ {     \lower.40ex\hbox{$\longrightarrow$}
                   \atop \raise.20ex\hbox{$\longleftarrow$}
                   }     $}  }

\def\pmb#1{\setbox0=\hbox{$#1$}%
\kern-.025em\copy0\kern-\wd0
\kern.05em\copy0\kern-\wd0
\kern-.025em\raise.0433em\box0 }

%
%
\font\fivebf=cmbx5
\font\sixbf=cmbx6
\font\sevenbf=cmbx7
\font\eightbf=cmbx8
\font\ninebf=cmbx9
\font\tenbf=cmbx10

\font\bfmone=cmbx10 scaled\magstep1

\font\sevenit=cmti7
\font\eightit=cmti8
\font\nineit=cmti9
\font\tenit=cmti10

\font\itmone=cmti10 scaled\magstep1

\font\fiverm=cmr5
\font\sixrm=cmr6
\font\sevenrm=cmr7
\font\eightrm=cmr8
\font\ninerm=cmr9
\font\tenrm=cmr10

\font\rmmone=cmr10 scaled\magstep1

\def\fontone{\def\rm{\fcm0\rmmone}%
  \textfont0=\rmmone \scriptfont0=\tenrm \scriptscriptfont0=\sevenrm
  \textfont1=\itmone \scriptfont1=\tenit \scriptscriptfont1=\sevenit
  \def\it{\fcm\itfcm\itmone}%
  \textfont\itfcm=\itmone
  \def\bf{\fcm\bffcm\bfmone}%
  \textfont\bffcm=\bfmone \scriptfont\bffcm=\tenbf
   \scriptscriptfont\bffcm=\sevenbf
  \tt \ttglue=.5em plus.25em minus.15em
  \normalbaselineskip=25pt
  \let\sc=\tenrm
  \let\big=\tenbig
  \setbox\strutbox=\hbox{\vrule height10.2pt depth4.2pt width\z@}%
  \normalbaselines\rm}



\font\ninerm=cmr9
\font\eightrm=cmr8
\font\sixrm=cmr6

\font\ninei=cmmi9
\font\eighti=cmmi8
\font\sixi=cmmi6
\skewchar\ninei='177 \skewchar\eighti='177 \skewchar\sixi='177

\font\ninesy=cmsy9
\font\eightsy=cmsy8
\font\sixsy=cmsy6
\skewchar\ninesy='60 \skewchar\eightsy='60 \skewchar\sixsy='60

\font\ninebf=cmbx9
\font\eightbf=cmbx8
\font\sixbf=cmbx6

\font\ninett=cmtt9
\font\eighttt=cmtt8

\hyphenchar\tentt=-1 
\hyphenchar\ninett=-1
\hyphenchar\eighttt=-1

\font\ninesl=cmsl9
\font\eightsl=cmsl8

\font\nineit=cmti9
\font\eightit=cmti8


\newskip\ttglue
\def\tenpoint{\def\rm{\fcm0\tenrm}%
  \textfont0=\tenrm \scriptfont0=\sevenrm \scriptscriptfont0=\fiverm
  \textfont1=\teni \scriptfont1=\seveni \scriptscriptfont1=\fivei
  \textfont2=\tensy \scriptfont2=\sevensy \scriptscriptfont2=\fivesy
  \textfont3=\tenex \scriptfont3=\tenex \scriptscriptfont3=\tenex
  \def\it{\fcm\itfcm\tenit}%
  \textfont\itfcm=\tenit
  \def\sl{\fcm\slfcm\tensl}%
  \textfont\slfcm=\tensl
  \def\bf{\fcm\bffcm\tenbf}%
  \textfont\bffcm=\tenbf \scriptfont\bffcm=\sevenbf
   \scriptscriptfont\bffcm=\fivebf
  \def\tt{\fcm\ttfcm\tentt}%
  \textfont\ttfcm=\tentt
  \tt \ttglue=.5em plus.25em minus.15em
  \normalbaselineskip=16pt
  \let\sc=\eightrm
  \let\big=\tenbig
  \setbox\strutbox=\hbox{\vrule height8.5pt depth3.5pt width\z@}%
  \normalbaselines\rm}

\def\ninepoint{\def\rm{\fcm0\ninerm}%
  \textfont0=\ninerm \scriptfont0=\sixrm \scriptscriptfont0=\fiverm
  \textfont1=\ninei \scriptfont1=\sixi \scriptscriptfont1=\fivei
  \textfont2=\ninesy \scriptfont2=\sixsy \scriptscriptfont2=\fivesy
  \textfont3=\tenex \scriptfont3=\tenex \scriptscriptfont3=\tenex
  \def\it{\fcm\itfcm\nineit}%
  \textfont\itfcm=\nineit
  \def\sl{\fcm\slfcm\ninesl}%
  \textfont\slfcm=\ninesl
  \def\bf{\fcm\bffcm\ninebf}%
  \textfont\bffcm=\ninebf \scriptfont\bffcm=\sixbf
   \scriptscriptfont\bffcm=\fivebf
  \def\tt{\fcm\ttfcm\ninett}%
  \textfont\ttfcm=\ninett
  \tt \ttglue=.5em plus.25em minus.15em
  \normalbaselineskip=11pt
  \let\sc=\sevenrm
  \let\big=\ninebig
  \setbox\strutbox=\hbox{\vrule height8pt depth3pt width\z@}%
  \normalbaselines\rm}

\def\eightpoint{\def\rm{\fcm0\eightrm}%
  \textfont0=\eightrm \scriptfont0=\sixrm \scriptscriptfont0=\fiverm
  \textfont1=\eighti \scriptfont1=\sixi \scriptscriptfont1=\fivei
  \textfont2=\eightsy \scriptfont2=\sixsy \scriptscriptfont2=\fivesy
  \textfont3=\tenex \scriptfont3=\tenex \scriptscriptfont3=\tenex
  \def\it{\fcm\itfcm\eightit}%
  \textfont\itfcm=\eightit
  \def\sl{\fcm\slfcm\eightsl}%
  \textfont\slfcm=\eightsl
  \def\bf{\fcm\bffcm\eightbf}%
  \textfont\bffcm=\eightbf \scriptfont\bffcm=\sixbf
   \scriptscriptfont\bffcm=\fivebf
  \def\tt{\fcm\ttfcm\eighttt}%
  \textfont\ttfcm=\eighttt
  \tt \ttglue=.5em plus.25em minus.15em
  \normalbaselineskip=9pt
  \let\sc=\sixrm
  \let\big=\eightbig
  \setbox\strutbox=\hbox{\vrule height7pt depth2pt width\z@}%
  \normalbaselines\rm}


\magnification=1200

\vsize=7.5in
\hsize=5.5in
\tolerance 10000

\baselineskip 12pt plus 1pt minus 1pt
\pageno=0
\centerline{\bf THE FUNCTIONAL INTEGRAL FOR A FREE}
\smallskip
\centerline{{\bf PARTICLE ON A HALF-PLANE}\footnote{*}{This
work is supported in part by funds
provided by the U. S. Department of Energy (D.O.E.) under contract
\#DE-AC02-76ER03069, and by the Fonds pour la Formation de Chercheurs et
l'Aide \`a la Recherche.}}
\vskip 24 pt
\centerline{Michel Carreau}
\vskip 12pt
\centerline{\it Boston University}
\centerline{\it Department of Physics}
\centerline{\it 590 Commonwealth Avenue}
\centerline{\it Boston, Massachusetts\ \ 02215\ \ \ U.S.A.}
\centerline{E-mail:CARREAU@BUPHY.BU.EDU}
\vskip 1.5in
\centerline{To be published in : {\it Journal of Mathematical Physics}}
\vfill
\centerline{ Typeset in $\TeX$ }
\vskip -12pt
\noindent BU-HEP-91-23\hfill November 1991
\eject

\centerline{\bf ABSTRACT}
\medskip
A free non-relativistic particle moving in two dimensions on a half-plane
can be described by self-adjoint Hamiltonians characterized by
boundary conditions imposed on the systems. The most general boundary condition
is parameterized in terms of the elements of an infinite-dimensional matrix.
We construct the Brownian functional
integral for each of these self-adjoint Hamiltonians.
Non-local boundary conditions are implemented by allowing the paths
striking the boundary to jump to other locations on the boundary.
Analytic continuation in time results in the Green's
functions of the Schr\"odinger
equation satisfying the boundary condition characterizing the
self-adjoint Hamiltonian.
\vfill
\eject
\noindent{\bf I.\quad INTRODUCTION}
\medskip
\nobreak
In a paper by Clark et al., and later  on by Gaveau et al. and Farhi et al.
(see Ref.~[1]) the functional integral for a  particle moving on a half-line
was derived.  It was found that the measure for the
functional integral  depends upon the one parameter family of boundary
conditions
which guarantee that probability does not not leave the half-line.
Farhi, Gutmann and the present author$^2$ derived the functional integral
for a free particle moving in a box in 1-D for the system subject to a four-
parameter family of boundary conditions consistent with unitarity.
They found that the measure for the
functional integral depends upon these four parameters and that the
paths are allowed to jump from wall to wall.

In this paper, we show how to generalize these ideas to more than one
dimension.  In Section~II, we describe the most general boundary condition
leading to unitary time evolution for the free particle
moving on a half-plane.  (Equivalently, we find the self-adjoint extensions of
the free Hamiltonian.)
 In Section~III, we construct the Brownian functional integral representation
of the Brownian Green's function corresponding to each self-adjoint
Hamiltonian.  The Green's function of the Schr\"odinger equation are obtained
by an analytic continuation of the Brownian Green's function.
We point out how to generalize these results to more complex quantum systems.
In the Appendix we give the functional integrals of a one dimensional quantum
systems made out of any number of half-lines joined together at their origin.
\goodbreak
\bigskip
\noindent{\bf II.\quad PARTICLE ON A HALF-PLANE --- SELF-ADJOINT EXTENSIONS}
\medskip
\nobreak
Consider the motion of a free particle on a half-plane in two dimensions
(${-\infty} <y<{+\infty}$
and $x>0$).  The Hilbert space consists of all square integrable functions on
the half-plane. The domain of the Hamiltonian
$$\hat H = - {1\over 2}\ {d^2 \over dx^2}- {1\over 2}\ {d^2 \over dy^2}
\eqno(2.1)$$
must be chosen so that $\hat H$ is Hermitian,
which is the condition
$$\int^{+\infty}_{-\infty} \int^\infty_0 \psi^* \left(\hat H\phi\right)
dx\,dy = \int^{+\infty}_{-\infty} \int^\infty_0 \left(\hat H\psi\right)^*
\phi\, dx\,dy \eqno(2.2)$$
or equivalently
$$\int^{+\infty}_{-\infty} dy\left[ \psi^* {d\phi\over dx} - {d\psi^*\over
dx}\phi\right]\Bigg|_{x=0} = 0 \eqno(2.3)$$
for all $\psi$ and $\phi$ in the domain of $\hat H$.

The condition (2.3) can be realized most generally if for every
$\chi$ in the domain of $\hat H$
$${d\chi\over dx}(x,y) \bigg|_{x=0} = \int^{+\infty}_{-\infty} dy' g(y',y)
\chi(0,y')\eqno(2.4)$$
for each $y$, with $g$ an arbitrary function obeying
$$g(y',y) = g^* (y,y')\ \ .\eqno(2.5)$$
A similar result was obtained in Ref.~[3].
Let us pick the following parameterization for $g(y',y)$
$$g(y',y) = \left\{ \left[ \rho(y) + \alpha(y)\right] \delta(y'-y) -
\rho(y')f(y',y)\,e^{ih(y',y)} \right\} \eqno(2.6)$$
where $\delta$ is the delta-function and $\alpha$, $h$, $\rho,f$ are
arbitrary smooth real functions with the following
restrictions: $\rho,f$ are non-negative, $h$ is antisymmetric in
its arguments,
$$\rho(y) f(y,y') = \rho(y') f(y',y)\eqno(2.7)$$
and
$$\int^{+\infty}_{-\infty} dy\,f(y',y) = 1\ \ .\eqno(2.8)$$
  We can show that $g(y',y)$ uniquely determines the functions
$\alpha$, $h$, $\rho,f$.
Given a $g$ satisfying (2.5) if we choose the domain of $\hat H$ to be
functions obeying (2.4) then
$\hat H$ is self-adjoint and probability does not leave the half-plane.  We
see that we have infinitely many self-adjoint extensions indexed by the
function
$g(y',y)$.
To each self-adjoint extension there corresponds a unique
Green's
function $G_g ({\svec z}_i,{\svec z}_f,t)$ with ${\svec z}_a = (x_a,y_a)$,
$a = i,f$ on the half-plane. The
Green's function can be expressed as
$$G_g ({\svec z}_i, {\svec z}_f,t) = \langle {\svec z}_f| e^{-i\hat H t}|
{\svec z}_i\rangle_g \ \ .\eqno(2.9)$$
Thought of as
a function of ${\svec z}_f$, $G_g$ satisfies the Schr\"odinger equation and the
boundary condition (2.4), that is
$${d\over dx_f}G_g({\svec z}_i,{\svec z}_f,t)\bigg|_{x_f=0}
 = \int dy'_f\,g(y_f',y_f)G_g({\svec z}_i,(0,y_f'),t)\ \ .\eqno(2.10)$$
If the boundary condition (2.4) has the property that when it is satisfied by
$\chi(x,y)$ it is also satisfied by $\chi(x,y+y_0)$ for an arbitrary $y_0$, we
say it is translationally invariant in the $y$-direction.  If we want a
translationally invariant boundary condition in the $y$-direction, then
$g(y,y')$ can depend only on $y-y'$.

Condition (2.4) says that the {\it net\/} current flowing out of the
half-plane is zero.  The condition that the current flow at each point on the
edge of the half-plane be zero is that the integrand in (2.3) vanishes for each
$y$.  This can be realized by taking $\rho(y) = 0$, leading to
$$g(y',y) = \alpha(y)\delta(y'-y)\ \ ,\eqno(2.11)$$
the most general {\it local\/} boundary condition.  If we want to enforce the
translation invariance in the $y$-direction
 and the condition that no current flow at
each $y$, we take $\alpha(y)$ independent of $y$. Then $g$ is simply
$$g(y',y) = \alpha\delta(y'-y)\ \ .\eqno(2.12)$$
With these restrictions there is only a one-parameter family of self-adjoint
extensions.
\goodbreak
\bigskip
\noindent{\bf III.\quad PARTICLE ON A HALF-PLANE --- FUNCTIONAL INTEGRAL}
\medskip
\nobreak
In this section, we construct the Brownian functional integrals corresponding
to the self-adjoint extensions of the free Hamiltonian constructed in the
previous section.  The analytic continuation of the Brownian functional
integrals results in the Green's functions of the Schr\"odinger equation.
\goodbreak
\bigskip
\noindent{\bf 1.\quad Preliminaries}
\medskip
\nobreak
In order to have a mathematically well-defined measure on a class of paths, we
work with Brownian functional integrals.  On the whole plane the Brownian
functional integral
$${\cal A}^{\rm B}_{\rm wp} \left( {\svec z}_i, {\svec z}_f,\tau\right) = \int
\left[ d{\svec q}_{\rm wp}\right]\,e^{- {1\over 2} \int^\tau_0
\dot{\svec q}^2_{\rm
wp}
\,d\tau'}
= {1\over 2\pi\tau} \,e^{- \left| {\svec z}_f - {\svec z}_i \right|^2/2\tau}
\eqno(3.1)$$
with ${\svec q}_{\rm wp}(0) = {\svec z}_i$ and ${\svec q}_{\rm wp}(\tau)
= {\svec z}_f$, is
the Green's function of the Schr\"odinger equation with imaginary time $ t =
-i\tau$.  We call ${\cal A}^{\rm B}_{\rm wp}$ the Brownian Green's function.
The symbol $[d{\svec q}_{\rm wp}]\,e^{-{1\over 2} \int^\tau_0
\dot{\svec q}^2_{\rm wp}\,d\tau'}$
is to be interpreted as the Brownian measure on paths ${\svec q}_{\rm wp}$.
This
measure gives zero weight to discontinuous paths.  The Green's function,
$G_{\rm wp}$, of the Schr\"odinger equation on the whole plane
is obtained by the analytic
continuation of ${\cal A}^{\rm B}_{\rm wp}$:
$$G_{\rm wp} \left( {\svec z}_i, {\svec z}_f,t\right) \equiv
{\cal A}^{\rm B}_{\rm wp} \left(
{\svec z}_i, {\svec z}_f, it\right)\ \ . \eqno(3.2)$$
When working on the half-plane, we use the {\it reflecting measure\/}, which is
induced by taking the absolute value of the $x$-component of whole plane
Brownian motion.  That is, if ${\svec q}_{\rm wp}= (q_{1{\rm wp}},
q_{2{\rm wp}})$ is a whole plane path then ${\svec q}^* \equiv\left(
|q_{1{\rm wp}}|, q_{2{\rm wp}}\right) \equiv (q^*_x,q_y)$ lies on the
half-plane and the probability distribution, or measure, on ${\svec q}_{\rm
wp}$ directly induces a probability distribution on ${\svec q}^*$.  For each
path ${\svec q}^* = (q^*_x,q_y)$ which goes from ${\svec z}_i$ to ${\svec
z}_f$ on the half-plane in time $\tau$, there exists a ``local time'' which
tells us how much time the path spends near the boundary of the half-plane
$(x\approx 0)$.
$$\ell(\tau) \equiv\lim\limits_{\epsilon\to 0} {1\over 2\epsilon} \int^\tau_0
I\left( q^*_x(\tau')<\epsilon\right) d\tau' \eqno(3.3)$$
where $I(A) = 1$ or 0 according to whether $A$ occurs or not.  (More
informally, we can think of $\ell(\tau)$ as $\int^\tau_0 \delta\left(
q^*_x(\tau')\right) d\tau'$.)

We now define the half-plane Brownian functional integral,
$${\cal A}^{\rm B}_r  \left( {\svec z}_i,{\svec z}_f,\tau\right) = \int
[d{\svec
q}^*]\,e^{-{1\over 2}\int^\tau_0 \dot{\svec
q}^{*2}\,d\tau'}\,e^{-r \ell(\tau)} \eqno(3.4)$$
where $r$ is a real parameter and the functional integral is over all
half-plane reflected paths ${\svec q}^* = (q^*_x,q_y)$ from ${\svec z}_i$ to
${\svec z}_f$ in time $\tau$.  The $[d{\svec q}^*]$ times the first
exponential is to be interpreted as the reflected Brownian measure. There is
a Brownian motion theorem$^4$ that the limit in (3.3) exists
for almost every path so (3.4) is sensibly defined.  The Brownian motion on
the half-plane, ${\svec
q}^*(\tau')$, is the Cartesian product of a reflecting Brownian motion in the
$x$-direction, $q^*_x(\tau')$, and a whole line Brownian motion in the
$y$-direction, $q_y(\tau')$.  Since these motions are completely
independent, we
can write the reflecting measure on the half-plane as the product of the
reflecting measure in the $x$-direction times the whole line measure in the
$y$-direction.  Moreover, the local time, $\ell(\tau)$, depends only on the
motion
in the $x$-direction. Therefore, the functional integral (3.4) factors as
follows
$$\eqalign{{\cal A}^{\rm B}_r \left( {\svec z}_i, {\svec z}_f,\tau\right) &=
{\cal A}^{\rm B}_{x,r} \left( x_i, x_f,\tau\right)
{\cal A}^{\rm B}_y (y_i, y_f,\tau) \cr
&\equiv\left\{ \int \left[ dq^*_x\right]\,e^{-{1\over 2}\int^\tau_0 \dot
q^{*2}_x\,d\tau'}\,e^{-r\ell(\tau)}\right\} \left\{ \int \left[
dq_y\right]\,e^{-{1\over 2} \int^\tau_0 \dot q^2_y\,d\tau'}\right\}
\cr}\eqno(3.5)$$
where ${\svec z}_a = (x_a, y_a)$, for $a = i,f$.  It follows that ${\cal
A}^{\rm B}_r$ satisfies the boundary condition
$${d\over dx_f}{\cal A}^{\rm B}_r \left( {\svec z}_i, {\svec z}_f,\tau\right)
\bigg|_{x_f=0}
= r
 {\cal A}^{\rm B}_r \left( {\svec z}_i, {\svec z}_f,\tau\right)
\bigg|_{x_f=0}\eqno(3.6)$$
since $A^{\rm B}_{x,r}$ has been shown$^1$ to satisfy the same boundary
condition.  The analytic continuation, $G_r \left( {\svec z}_i,{\svec z}_f,
t\right) \equiv {\cal A}^{\rm B}_r \left( {\svec z}_i, {\svec z}_f, it\right)$
is the Green's function of the Schr\"odinger equation for a free particle on
the half-plane satisfying (3.6).  This one-parameter family of Green's
functions corresponds to the one-parameter family of self-adjoint Hamiltonians
defined in the last section by (2.12), i.e. those with translationally
invariant local boundary conditions.
\goodbreak
\bigskip
\noindent{\bf 2.\quad General Local Boundary Condition}
\medskip
\nobreak
In this subsection, we build the Brownian functional integrals for the
self-adjoint
extensions corresponding to the boundary condition (2.4) with $g$ given by
(2.11).  Let us generalize
the term, $r \ell(\tau)$, in the functional integral (3.4) to
$$\int^\tau_0 \alpha \left( q_y (\tau')\right) \dot\ell (\tau')
d\tau'\eqno(3.7)$$
for a path ${\svec q}^* = (q^*_x,q_y)$ with local time $\ell$, where $\dot\ell
\equiv d\ell/d\tau'$ and $\alpha$ is an arbitrary  real function.  The factor
$\dot\ell$
gives a contribution different from zero whenever the paths hit the edge of
the half-plane.  More informally, we can think of (3.7) as
$$\int^\tau_0\int^{+\infty}_{-\infty} \alpha(y_0)
\delta\left(q_y(\tau') -
y_0\right) \delta\left( q^*_x (\tau')\right) dy_0\,d\tau'\ \ ,\eqno(3.8)$$
a continuous sum of delta-function potentials along the edge of the
half-plane with different strengths for different $y$. The functional integral
on the half-plane becomes
$${\cal A}^{\rm B}_\alpha \left( {\svec z}_i, {\svec z}_f,\tau\right) = \int
\left[ d{\svec q}^*\right]\,e^{-{1\over 2}\int^\tau_0 \dot{\svec q}^{*2}
d\tau' -
\int^\tau_0 \alpha\left( q_y (\tau')\right)\dot\ell(\tau') d\tau'} \ \
.\eqno(3.9)$$
The expression (3.7) exists for almost every path$^{5,6}$ so (3.9) is sensibly
defined.  In the next subsection we will show that the analytic continuation
$G_\alpha\left( {\svec z}_i,{\svec z}_f,t\right) \equiv A^{\rm
B}_\alpha\left( {\svec z}_i, {\svec z}_f,it\right)$ is the Green's function of
the Schr\"odinger equation satisfying the most general local boundary condition
$${d\over dx_f} G_\alpha\left( {\svec z}_i,
{\svec z}_f,t\right)\bigg|_{x_f=0} =
\alpha(y_f) G_\alpha \left( {\svec z}_i, {\svec z}_f,t\right)\bigg|_{x_f=0}\ \
,\eqno(3.10)$$
for each $y_f$. Let us mention that path integral representations of
Green's functions for several types of differential equations satisfying
local boundary conditions have already been studied extensively in more than
one dimension(see Ref.~[7] and references therein).  The equation (3.9) was
obtained in Ref.~[7] for the diffusion equation.  In the next section,
we generalize this result for non-local boundary conditions.
\goodbreak
\bigskip
\noindent{\bf 3.\quad The Functional Integral For General Boundary Condition}
\medskip
\nobreak
In this subsection, we construct a family of functional integrals
${\cal A}^{\rm B}_g$ parameterized with the function $g$, introduced in
Section~II, which satisfies the boundary condition
$${d\over dx_f} {\cal A}^{\rm B}_g ({\svec z}_i,{\svec z}_f,\tau)\bigg|_{x_f=0}
 = \int dy'_f
\,g(y_f',y_f) {\cal A}^{\rm B}_g ({\svec z}_i,(0,y_f'),\tau)
 \eqno(3.11)$$
The  analytic continuation of ${\cal A}^{\rm B}_g$ with respect to $\tau$
results in the Green's functions (2.9)
of the Schr\"odinger equation satisfying the non-local boundary
condition (2.4) or equivalently (2.10).

In order to obtain a functional integral satisfying the non-local boundary
conditions (3.11), we wish to define a measure giving non-zero weight to paths
that can jump to other points on the edge of the half-plane when they hit
the edge of the half-plane.
Moreover, this measure should depends on the four functions
$\rho,\alpha,f,h$, introduced in Section~II to parameterized the function
$g$, since the boundary condition (3.11) depends on them.
We will see below that when a path hit the edge of the half-plane at $y$
along y-axis,
$\rho(y)$ gives the rate at which the path jumps from that point and $f(y,y')$
gives the density of probability of jumping to $y'$ given it has jumped from
$y$.  The functions $\alpha(y)$ and $h(y,y')$ control the size of weights and
phases given to the path when it spends time near $y$ and when it jumps
from $y$ to $y'$ respectively.

We will see that the probability of jumping from a point
on the edge of the half-plane depends both on the time the path spends
near that point and on the rate of jumping at that point.
It is then natural to introduce the notion of {\it effective local time\/}
which is roughly the amount of time the path spends near the edge of the
half-plane weighted, point-by-point, by the rate at which the path jumps
at that point. This notion is defined formally by (3.16) below.

Let us describe the process generating the paths on the half-plane.
Imagine that the path starts at ${\svec z}_i\equiv(x_i,y_i)$, with $x_i>0$ and
undergoes ordinary reflected Brownian motion on the half-plane until its
effective local time(defined below) reaches some number $s_1$.
(At the instant the effective local time reaches $s_1$
the path must be on the edge of the half-plane, say $(0,y^-_{1})$.)  It
then jumps
to $(0,y^+_{1})$ on the edge of the half-plane.  From this point it
resumes its reflected Brownian
motion on the half-plane until the effective local time reaches $s_1+s_2$
at which time it reaches $(0,y^-_{2})$ and jumps to $(0,y^+_{2})$, and
so forth.

The values $s_1$, $s_2$, $\ldots$ are to be chosen
as independent random variables which are exponentially distributed i.e.
$$\hbox{Prob} \left(s_{k}>u\right) = e^{-u}\ \ .\eqno(3.12)$$
As we will see later, the rate at which the paths jump is implicit in the
definition of the effective local time, (3.16), and this is why we have set the
coefficient of $-u$ to be $1$ without lost of generality.
The values $y^+_1$, $y^+_2$, $\ldots$  are to be chosen as random
variables and their distribution depends respectively on
the locations $(0,y^-_{1})$, $(0,y^-_{2})$, $\ldots$,
on the edge of the half-plane, from
which the jump takes place, that is
$$\hbox{Prob}\left(w<y^+_{k}<w+dw\right) =
f (y^-_{k}, w)\,dw\ \ .\eqno(3.13)$$
We see that $f(y^-_k,w)$ is the probability density of
landing at $(0,w)$ given that the path has jumped from $(0,y^-_k)$.

More formally, let ${\svec q}^* = (q^*_x,q_y)$ be an ordinary reflected
Brownian motion on the half-plane as described in Subsection~III.1 and let
$\ell$ be its local time defined via (3.3).  Given the values
$s_1$, $y^+_1$, $s_2$, $y^+_2$, $\ldots$ as described above,
we define the path ${\svec q}$ by
$${\svec q}(\tau')\equiv \left( q_1(\tau'), q_2(\tau')\right)
\equiv\left( q^*_x(\tau'), q_y (\tau') +\Delta y_1 + \Delta y_2 +
\ldots +\Delta y_\#\right)\eqno(3.14)$$
where $\Delta y_k\equiv y^+_k-y^-_k$, $\#$ is the number of jump,
$$s_1+s_2 + \ldots +s_\# \le s(\tau') <s_1 + \ldots + s_{\#+1}\eqno(3.15)$$
and
$$s(\tau')\equiv \int^{\tau'}_{0} \rho\left(q_2(\tilde\tau)
\right)\,\dot{\ell}(\tilde\tau)\,d\tilde\tau\ \ .\eqno(3.16)$$
Here, $\rho(y)$ is the rate per unit local time at which the path jumps
from the point $(0,y)$ on the edge of the half-plane.
We call $s(\tau')$ the effective local time.  It measures the
``time'' the path spends near the edge of the half-plane
weighted, point-by-point, by the rate $\rho(y)$ at which the path
jump at that point. Definitions (3.14) and
(3.16) require further explanation since they are defined in terms of each
other.  Let (3.14) be given for the first $\#$-jumps of the path, then
(3.16) is well defined until
$s(\tau')$ reaches $s_1+s_2+\ldots+s_{\#+1}$.  At this time the path is
at $(0,y^-_{\#+1})$ and jumps to $(0,y^+_{\#+1})$. This determines
$\Delta y_{\#+1}$
in definition (3.14), and (3.16) is updated to take this jump into
account. $s(\tau')$ is now well defined until it reaches
$s_1+s_2+\ldots+s_{\#+2}$... and so forth.  The above is true for
$\#=0,1,2,\ldots$ so definitions (3.14) and (3.16) should now be clear.
Note that the values $y^-_1$, $y^-_2$, $\ldots$ are completely
determined by the
half-plane reflected path ${\svec q}^*$ and the values $s_1$, $y^+_1$, $s_2$,
$y^+_2$, $\ldots$.

Let $\tau_1$, $\tau_2$, $\ldots$, $\tau_\#$ be the times at which the effective
local time reaches $s_1$, $s_1+s_2$, $\ldots$
$s_1+s_2+\ldots +s_\#$.
We note that $q_2(\tau')$ is discontinuous at $\tau_1$, $\tau_2$, $\ldots$,
$\tau_\#$. The positions of the path along the $y$-direction at time $\tau_k$,
before and after the $k$-th jump can be expressed in terms of
${\svec q}=(q_1,q_2)$
as
$$y^\pm_{k} \equiv\lim\limits_{\epsilon\to 0} q_2 (\tau_k \pm\epsilon)\ \ .
\eqno(3.17)$$


The measure on paths ${\svec q}$ , denoted as $[d{\svec q}]_{\rho,f}
\,e^{-{1\over 2} \int^\tau_0 \dot{\svec q}^2 d\tau'}$, is induced from the
measure on half-plane reflected paths and the distribution (3.12) and (3.13).
We use this measure to define the functional integral
$${\cal A}^{\rm B}_g \left( {\svec z}_i, {\svec z}_f,\tau\right) = \int
[d{\svec q}]_{\rho,f}\,e^{- {1\over 2}\int^\tau_0 \dot{\svec q}^2 d\tau'}\,e^{-
\int^\tau_0 \alpha\left( q_2(\tau')\right)\dot\ell (\tau')d\tau'}
e^{i\sum\limits^{\#}_{k=1}
h \left( y^-_{k}, y^+_{k}\right)} \
. \eqno(3.18)$$
The functional integral is over all paths ${\svec q}$ from
${\svec z}_i\equiv (x_i,y_i)$ to ${\svec z}_f\equiv (x_f,y_f)$ in time
$\tau$.
In (3.18), $\rho$, $\alpha$, $h$, $f$ are the four functions introduced in
Section~II in order to parameterize the function $g$ specifying the boundary
condition (2.4). \# is the number of jumps of a given path $\svec q$.  The
last factor in (3.18), involving $h$, is a product of phases where each factor
is the phase associated with one jump of the path. The arguments of $h$
represent the location from and to which the path jump respectively along
the $y$-axis.
The factor involving $\alpha$ was discussed in the previous subsection.

The memoryless property of the distribution (3.12) and the fact that
the distribution of $y^+_k$, (3.13),
does not depend on past events but only on the present position of the path
before a
jump guarantees that ${\cal A}^{\rm B}_g$ satisfies the convolution equation
$${\cal A}^{\rm B}_g \left( {\svec z}_i, {\svec z}_f, \tau_1+\tau_2\right) =
\int\limits_{{\rm half-}\atop{\rm plane}} d^2 {\svec z}\,{\cal A}^{\rm B}_g
\left( {\svec z}_i, {\svec z},\tau_1\right) {\cal A}^{\rm B}_g \left(
{\svec z}, {\svec z}_f,\tau_2\right) \eqno(3.19)$$
(To establish this, we used the fact that (3.7) and the last factor
in (3.18) convolve
path-by-path).
${\cal A}^{\rm B}_g$ also satisfies
$${\cal A}^{\rm B}_g \left( {\svec z}_i, {\svec z}_f,\tau\right) = {\cal
A}^{\rm B}_g \left( {\svec z}_f, {\svec z}_i,\tau\right)^* \eqno(3.20)$$
because for each path ${\svec q}$ from ${\svec z}_i$ to ${\svec z}_f$
there exists a path ${\svec q}'$ from ${\svec z}_f$ to ${\svec z}_i$
with weight complex conjugate to the weight of the path ${\svec q}$.
For ${\svec z}_i$ and ${\svec z}_f$ not on the edge of the
half-plane, and for $\tau$ small (so that paths which hit the boundary can be
neglected), ${\cal A}^{\rm B}_g$ is well-approximated by ${\cal A}^{\rm
B}_{\rm wp}$.  Combined with (3.19) this ensures that ${\cal A}^{\rm B}_{g}$
satisfies the same differential equation as ${\cal A}^{\rm B}_{\rm wp}$
for all $\tau$ and for all ${\svec z}_i,{\svec z}_f$ not on the edge of the
half-plane:
$$-{1\over 2}\left( {\partial^2\over\partial x^2_f} + {\partial^2\over\partial
y^2_f}\right) {\cal A}^{\rm B}_g \left( {\svec z}_i, {\svec z}_f,\tau\right)
= -
{d\over d\tau} {\cal A}^{\rm B}_g \left( {\svec z}_i, {\svec z}_f,\tau\right)
\eqno(3.21)$$
with
$${\cal A}^{\rm B}_g \left( {\svec z}_i, {\svec z}_f, 0 \right) = \delta\left(
{\svec z}_f - {\svec z}_i\right)\ \ .\eqno(3.22)$$
The fact that ${\cal A}^{\rm B}_g$ satisfies the last four equations is
sufficient to guarantee it satisfies some boundary condition of the form
(2.4). In the next subsection, we will show that
indeed ${\cal A}^{\rm B}_g$ satisfies the boundary condition (3.11).
The analytic continuation $G_g ({\svec z}_i,{\svec z}_f,t) \equiv
{\cal A}^{\rm B}_g({\svec z}_i, {\svec z}_f,it)$ satisfies (3.11) also.
So $G_g$ is the Green's function, (2.9),
of the Schr\"odinger equation for the
self-adjoint Hamiltonian corresponding to the boundary condition (2.4).

We can obtain the functional integral of more complicated quantum systems
as follows.
For a quantum system where the particle is limited to move inside a
bounded  region, one can show that the self-adjoint-extension of the
Hamiltonain are indexed by
a function $g$ defined on the boundary of the region. For each
self-adjoint-extension, we can construct the functional integral
in terms of functions $\rho,f,\alpha,h$,
defined on the boundary of the region, parameterizing  the function $g$.
The functional integral we get is very similar  to (3.18).
The reflected Brownian measure needed in the definition of the
the functional integral is mathematically well-defined for
region with smooth boundary and can be induced from the measure on whole
line path.

So far in this paper, we have only considered the motion of a free particle.
We can obtain the functional integral for a particle moving under the influence
of a bounded potential, $U({\svec q})$, by multiplying the
integrand of the functional integral (3.18) by
$$e^{-\int^\tau_0 U\left( {\svec q}(\tau')\right) d\tau'}\ \ .\eqno(3.23)$$
Both the self-adjoint extensions and the proof that the functional integral
satisfies the appropriate boundary condition are unaffected by the
introduction of that factor.
\goodbreak
\bigskip
\noindent{\bf 4.\quad Proof of Boundary Condition for Functional Integral}
\medskip
\nobreak
In the previous subsection, we showed that the functional integral
$${\cal A}^{\rm B}_g \left( {\svec z}_i, {\svec z}_f,\tau\right) = \int
[d{\svec q}]_{\rho,f}\,e^{- {1\over 2}\int^\tau_0 \dot{\svec q}^2 d\tau'}\,e^{-
\int^\tau_0 \alpha\left( q_2(\tau')\right)\dot\ell (\tau')d\tau'}
e^{i\sum\limits^{\#}_{k=1}
h \left( y^-_{k}, y^+_{k}\right)}\eqno(3.24)$$
satisfies some boundary condition of the form (2.4). In this subsection,
we prove that the functional integral satisfies the specific boundary condition
$${d\over dx_f} {\cal A}^{\rm B}_g ({\svec z}_i,{\svec z}_f,\tau)\bigg|_{x_f=0}
 = \int dy'_f
\,g(y_f',y_f) {\cal A}^{\rm B}_g ({\svec z}_i,(0,y_f'),\tau)\ \ .\eqno(3.25)$$
We note that this boundary condition, regarded as a function of
${\svec z}_f$, is the same for each $\left( {\svec z}_i,\tau\right)$.
Also, we remark that we only need to determine ${\cal A}^{\rm B}_g$ near the
edge of the half-plane and at a particular point along the y-axis in order
to distinguish which boundary condition it satisfies among the different
possibilities offered by (2.4). In other words, it is sufficient to analyze
${\cal A}^{\rm B}_g$ in the limit
$$\left. \eqalign{ \tau&\phantom{=} \hbox{small}\cr
{\svec z}_i &\equiv\left(x_i,y_i\right)=(0,y_i) \cr
{\svec z}_f &\equiv\left(x_f,y_f\right)\approx (0,y_f) \cr} \right\}
\eqno(3.26)$$
where $x_f$ is much smaller than $\sqrt{2\tau}$ and $y_f=y_i$. As the following
shows, in this limit the path integral (3.24) is well-approximated
by a sum over dominant paths. By Taylor expanding the last two factors
in (3.24) around the dominant paths we see that to leading orders
the Brownian motion in the y-direction decouples from the jumping process and
the Brownian motion in the x-direction. Once the Brownian motion in the
y-direction is integrated out, the path integral we are left with is
recognized to be a good approximation for the path integral of
a 1-dimensional quantum system made out of a continuous set of half-lines
all joined together at the origin(see the Appendix).
This identification enables us to use the techniques of Ref.~[2] to
show which boundary condition is satisfied by the path integral.

In the small $\tau$-limit, the main contribution to the functional
integral (3.24) comes from paths ${\svec q}(\tau')$ that remain within a circle
of radius $\sqrt{2\tau}$ centered at $(x_i, y_i)$ before the first jump
and $(x_i, y^+_k)$ between the $k$-th and $(k+1)$-th jump.
This is because between jumps the paths are generated by Brownian motion which
have an exponentially small probability of reaching a point at a distance
greater than $\sqrt{2\tau}$ from its starting point.
Moreover, in the small $\tau$-limit, paths for which $\ell(\tau)$ is
greater than $\sqrt{2\tau}$ are exponentially suppressed$^{2,6}$.
This implies, as shown in Ref.~[2],
that paths with $\#$-jumps are suppressed by a factor $(2\tau)^{\#/2}$.
One can show that paths jumping less than three times contain all of the
information necessary to determine which boundary condition is satisfied by
${\cal A}^{\rm B}_g$. To establish this, it is sufficient to  observe that,
as a whole, these paths are sensible to all the values of the functions
$\rho,f,\alpha,h$: that is, any point on the boundary can be visited by
at least
one of these paths. If we had neglected paths that jump twice, this statement
would not have been true. We can neglect paths that jump more
than twice.

To sum up, we have seen that the dominant paths
in the functional integral (3.24) jump less than three times, have a local
time less than $\sqrt{2\tau}$ and
remain within a circle of radius $\sqrt{2\tau}$ from the points
$(x_i, y^+_k)$. (Note that $y^+_0\equiv y_i$.) Also, we have seen that we can
drop any contributions to the functional integral suppressed by at least a
factor $(2\tau)^{3/2}$ since we have neglected paths with three jump and more.

In the functional integral (3.24) the functions $\rho,f,\alpha,h$
depend on the paths. Since a given
path $\svec q$, with $\#$-jump, deviates by at most $\sqrt{2\tau}$ from the
points $(x_i, y^+_k)$, we can perform a
Taylor expansion of $\rho,f,\alpha,h$ around these points. With a little
bit of work, we can show that we need only to keep the leading terms in the
expansion of these functions. For the contribution to the functional
integral coming from the next to leading terms different from zero
in this expansion are suppressed by a factor $(2\tau)^{3/2}$.
(for the case where $\rho,f,\alpha,h$ are very slowly varying
it is clear that it is sufficient to keep only the leading term on their
expansion.)

By expanding $\rho, f,\alpha,h$ to leading order,
we can rewrite the functional integral (3.24) as
$${\cal A}^{\rm B}_g \left( {\svec z}_i, {\svec z}_f,\tau\right)\approx
\int_{\cal D}
[d{\svec q}]_{\rho,f}\,e^{- {1\over 2}\int^\tau_0 \dot{\svec q}^2 d\tau'}\,
e^{-\sum\limits^{\#}_{k=0} \alpha(y^+_{k}) \ell_k}
e^{i\sum\limits^{\#-1}_{k=0}h\left( y^+_{k}, y^+_{k+1}\right)}\eqno(3.27)$$
where
$\cal D$ stands for the path integral over all the dominant
paths from ${\svec z}_i$ to ${\svec z}_f$; that is the paths
that remain within a circle of radius $\sqrt{2\tau}$ from the points
$(x_i, y^+_k)$, with zero, one
and two jumps($\#=0,1,2$) and with local time less than $\sqrt{2\tau}$.
(we have set $y^+_0\equiv y_i$ and $y^-_k\approx y^+_{k-1}$, for $k=1,2,
\ldots$.) Also,
$$\ell_k \equiv\lim\limits_{\epsilon\to 0} {1\over 2\epsilon}
\int^{\tau_{k+1}}_{\tau_k}
I\left( q_1(\tau')<\epsilon\right) d\tau' \eqno(3.28)$$
is the local time the path
spends near the origin between the $k$-th and $(k+1)$-th jump
for $k=1,2,\ldots,\# -1$. $\ell_0$ and $\ell_\#$
is the local time the  path spends near the origin before the first
and after the last jump respectively. Finally, the effective local
time in the functional integral is
$$s(\tau)\approx\sum^{\#}_{k=0} \rho(y^+_{k})\ell_k\eqno(3.29)$$
to leading order in the expansion of $\rho$.

Now, let us make two remarks. First recall that a path ${\svec q}$ can be
expressed in terms of a half-plane reflected path ${\svec q}^*=(q^*_x, q_y)$
from ${\svec z}_i$ to an intermediate point $(x_i, \tilde y)$ and in
terms of the values $s_1$, $y^+_1$, $s_2$, $y^+_2$, $\ldots$ restricted to
$$\sum\limits^{\#}_{k=1} \Delta y_k\equiv\sum\limits^{\#}_{k=1}
(y^+_k-y^-_k)=y_f-\tilde y\ \ .\eqno(3.30)$$
This restriction ensures that the path ${\svec q}$ ends at ${\svec z}_f$
as required.  The path integral can be broken into a few integrations.
We begin by integrating over the values $s_1$, $y^+_1$, $s_2$, $y^+_2$,
$\ldots$ restricted to (3.30) with ${\svec q}^*$  and $\tilde y$ fixed.
Next we integrate over paths
$q^*_x$, $q_y$ with $\tilde y$ fixed and finally integrate
over all $\tilde y$.
As a second remark we point out that,
in the functional integral (3.27),  by relaxing
the domain of integration to be
over all paths instead of only those describe by $\cal D$ we only
introduce errors suppressed by at least a factor $(2\tau)^{3/2}$.

We can now rewrite (3.27) as
$$\eqalign{{\cal A}^{\rm B}_g\left( {\svec z}_i, {\svec z}_f,\tau\right)
&\approx\int^{+\infty}_{-\infty} d\tilde y\,
{\cal A}^{\rm B}_{\rm free}(y_i, \tilde y, \tau)\,
{\cal A}_{y_i, y_i+(y_f-\tilde y)}(x_i,x_f,\tau)
\cr
&\approx{\cal A}_{y_i, y_f}(x_i,x_f,\tau)
\cr}
\eqno(3.31)$$
where
$${\cal A}^{\rm B}_{\rm free}(y_i,\tilde y,\tau)= {1\over \sqrt{2\pi\tau}}\,
e^{-(\tilde y - y_i )^2/2\tau}\eqno(3.32)$$
is the result of the integration over all the paths $q_y$
(note that to leading order nothing but
$e^{- {1\over 2}\int^\tau_0 \dot{\svec q}^2 d\tau'}$depends on ${\svec q}_y$
in (3.27).),
and ${\cal A}_{y_i,y_f}$ is the result of the integration
over the values $s_1$, $y^+_1$, $s_2$, $y^+_2$, $\ldots$
restricted to (3.30) and over all paths $q^*_x$. In the appendix
we show that ${\cal A}_{y_i,y_f}$  can be thought of as
the functional integral of a 1-dimensional system made out of a
continuous set of
half-lines all joined together at the origin.  This system is a straightforward
generalization of a quantum system with only two half-lines connected
at the origin and this system was studied in a previous paper$^2$.
Using the same techniques introduced in Ref.~[2] we can show that
${\cal A}_{y_i,y_f}$ satisfies
$${d\over dx_f}{\cal A}_{y_i,y_f}(x_i,x_f)\bigg|_{x_f=0}
= \int^{+\infty}_{-\infty} dy_f' g(y_f',y_f){\cal A}_{y_i,y_f'}(x_i,0)
\eqno(3.33)$$
and in turn that implies that ${\cal A}^{\rm B}_g$ satisfies (3.25)
as we wanted to show.
\goodbreak
\bigskip
\noindent{\bf IV.\quad CONCLUSIONS}
\medskip
\nobreak
In this paper, the equivalence between the operator and functional approaches
to quantum mechanics established for one-dimensional quantum systems in
Ref.~[1,2,3] has been extended to quantum systems in higher dimension admitting
general boundary conditions.  For each self-adjoint extension the
Green's function is obtained as the analytic continuation of the functional
integral whose measure depends explicitly upon the parameters defining the
extension.
\goodbreak
\bigskip
\centerline{\bf ACKNOWLEDGMENTS}
\medskip
I am grateful to Edward Farhi and to Sam Gutmann for their supports,
insights and valuable remarks throughout the course of this
work.  I wish to thank Daniel Stroock, Barrett Rogers, Martin Leblanc and
the referee for helpful discussions and comments.
\par
\vfill
\eject
\centerline{\bf REFERENCES}
\medskip
\item{1.}T.E. Clark, R. Menikoff and D. H. Sharpe, {\it Phys. Rev.\/} {\bf
D22} (1980) 3012.
\medskip
E. Farhi and S. Gutmann,  {\it Int. Jour. Mod. Phys.} {\bf A5}
(1990) 3029.
\medskip
B. Gaveau and L. S. Shulman, {\it J. Phys. A.:\ Math. Gen.\/} {\bf
19} (1986) 1833.
\medskip
\item{2.}M. Carreau, E. Farhi, S. Gutmann, {\it Phys. Rev.} {\bf D42}
(1990) 1194.
\medskip
\item{3.}M. Carreau, E. Farhi, S. Gutmann and P. Mende,
{\it Ann. of Phys.} {\bf 1},(1990) 186.
\medskip
\item{4.}K. It\^o and H. McKean, Jr., {\it Diffusion Processes and Their
Sample Paths\/} (Springer-Verlag, New York, 1964), p.~63.
\medskip
\item{5.}Daniel Revuz, Marc Yor, {\it Continuous Martingales and
Brownian Motion\/} (Springer-Verlag, New York, 1991) Chap. VI.
\medskip
\item{6.}K. Sato and H. Tanaka, {\it Proc. Japan Acad.} {\bf 38} (1962) 699
\medskip
\item{7.}K. Sato and T. Ueno, {\it J. Math. Kyoto Univ.} {\bf 4}
(1965) 529
\medskip
N. Ikeda, {\it Mem. Coll. Sci. Univ. Kyoto, Ser. A,} {\bf 33} (1961) 367
\medskip
\goodbreak
\bigskip
\vfill
\eject
\noindent{\bf APPENDIX}
\medskip
\nobreak
Consider the quantum system made out of N-half-lines
parameterized by $k=1,2,\ldots, N$ all joined together at the origin
as indicated in Fig.1.  We locate the position of a particle on this
system by specifying the branch, $k$, where the particle moves, and its
distance, $x$, from the origin. The wave function $\chi_k(x)$ for each branch
$k$ of this system satisfies the
Schr\"odinger equation.
In order to make the Hamiltonian self-adjoint the wave function
must satisfy the boundary condition
$${d\over dx}{\chi_j}(x)\bigg|_{x=0}
= \sum\limits^{N}_{k=1}M_{k,j}\chi_k(0)\eqno(A.1)$$
where $M$ is a given hermitian matrix, characterizing a particular
self-adjoint extension of this system. We can parameterize $M$ in terms of
the  real parameters $\rho_j,\alpha_j, f_j, h_j$ as follows
$$M_{k,j}=(\rho_j +\alpha_j)\delta_{k,j}-{\rho_k}f_{k,j}e^{ih_{k,j}}
\ \ .\eqno(A.2)$$
where $\rho_k\ge 0$, $f_{k,j}\ge 0$, $f_{k,k}=0$,
$h_{k,j}=-h_{j,k}$, ${\rho_k}f_{k,j}={\rho_j}f_{j,k}$ and
$$\sum\limits^{N}_{j=1}f_{k,j}=1\ \ .\eqno(A.3)$$
Below we will use these functions to construct the functional
integral.

Let us describe the process generating the paths on this system.
The path starts on branch $i$ at $x_i$ and undergoes
reflected Brownian motion on this branch until its effective local
time(defined below) reaches $s_1$.  At this time the path is at the
origin and it jumps to branch $i'$ at its origin and the path resumes
its reflected Brownian motion until
its effective local time reaches $s_2$ and so forth.  The values $s_1$,
$s_2$, $\ldots$  are chosen to be independent random variables which
are exponentially distributed, i.e.
$$\hbox{Prob} \left(s_{k}>u\right) = e^{-u}\ \ .\eqno(A.4)$$
Let $k_0$, $k_1$, $\ldots$, $k_\#$ be the branch in which the path is moving
before the first jump, after the first, second and last jump respectively.
We choose the values $k_1$, $k_2$, $\ldots$ to be random
variables with the following  distribution (note that $k_0\equiv i$
is the branch in which the path starts its motion.)
Given that the path has jumped from branch
$k_j$, the distribution of the value $k_{j+1}$ is
$$\hbox{Prob}\left(k_{j+1}=k'\right) = f_{k_j,k'}\ \ .\eqno(A.5)$$
We see that $f_{k_j,k'}$ is the probability of jumping to branch $k'$
given that the path had jumped from branch $k_j$.
Let $\ell_0$, $\ell_1$, $\ldots$, $\ell_\#$ be the local
time the path spends near the origin in the branch $k_0$ before the
first jump, and on branch $k_1$, $k_2$, $\ldots$, $k_\#$ after the first,
second, ..., last jump respectively,
we define the effective local time as
$$s(\tau')=\sum^{\#}_{j=0} \rho_{k_j}\ell_j(\tau')\eqno(A.6)$$
where $\#$ is the number of jumps of a given path and
$\rho_{k_j}$ is the rate per unit local time at which the paths jump from
the origin of branch $k_j$.

More formally let $q^*_x$ be a 1-dimensional
reflected Brownian motion on the positive half-line.  With
$s_1$, $k_1$, $s_2$, $k_2$,  $\ldots$ as described above, we
define the path $\bar q$ by
$$\bar q(\tau')=\bigg\{q^*_x(\tau')\ \ {\rm ;\ \ in\ \ branch\ \ }k_{j}\bigg\}
\eqno(A.7)$$
where
$$s_1+s_2 + \ldots +s_j \le s(\tau') <s_1 + \ldots + s_{j+1}\ \ .\eqno(A.8)$$
The distribution $(A.4)$ and $(A.5)$  and
the reflected Brownian measure for each path $q^*_x$
induces a measure for each path $\bar q$.
The functional integral for this system is
$${\cal A}^{\rm B}_{i,j}\left(x_i, x_f,\tau\right) = \int
[d\bar q]_{\rho,f}\,e^{- {1\over 2}\int^\tau_0 \dot{\bar q}^2 d\tau'}\,
e^{-\sum\limits^{\#}_{j=0} \alpha_{k_j}\ell_j}
e^{i\sum\limits^{\#-1}_{j=0}h\left( k_j, k_{j+1}\right)}\eqno(A.9)$$
where the path integral is over all paths $\bar q$ from $x_i$
on branch $i$ to $x_f$ on branch $j$ in time $\tau$.
Note that $k_0\equiv i$, $k_\#\equiv j$ and that
$e^{ih_{k,k'}}$ is a phase given to  the path whenever it jumps
from branch $k$ to $k'$.

Equivalently the path integral can be perform by first integrating over
all values $s_1$, $k_1$, $s_2$,  $k_2$, $\ldots$ restricted to
$$\sum\limits^{\#}_{j=1} (k_j-k_{j-1})\, mod\, N=j-i\eqno(A.10)$$
with $q^*_x$ fixed and then integrating over all paths $q^*_x$.

The functional integral $(A.9)$ satisfies
$${d\over dx_f}{\cal A}^{\rm B}_{i,j}(x_i,x_f)\bigg|_{x_f=0}
= \sum\limits^{N}_{k=1}M_{k,j}{\cal A}^{\rm B}_{i,k}(x_i,0)
\eqno(A.11)$$
where $M$ is given by $(A.2)$.
The proof of $(A.11)$ is a straight forward generalization
of the proof of the corresponding result for $N=2$ given
in Ref.~[1] and will not be repeated here:
for $N=2$, $(A.11)$ reduces to the Eq.(3.22) of Ref.~[1] since
hermiticity of the matrix $M$ implies that
$\rho_1=\rho_2$, $f_{1,2}=f_{2,1}=1$ and $h_{1,2}\equiv\theta$.

Now consider the quantum system made out of an infinite number of half-line
parameterized by $-\infty<y<\infty$ all joined together at the origin.
The path on this system starts at $x_i$ on branch $y_0$
jumps to branch $y_1$, $y_2$, $\ldots$, $y_\#$ whenever its effective local
time reaches $s_1$, $s_2$, $\ldots$, $s_\#$ respectively.  We denote
as $\ell_0$, $\ell_1$, $\ldots$, $\ell_\#$ the local time the path spends
near the origin before the first jump, after the second, third, ..., last
jump respectively.   The effective local time is defined as
$$s(\tau')=\sum^{\#}_{k=0} \rho(y_{k}) \ell_k(\tau')\eqno(A.12)$$
where $\#$ is the number of jumps of a given path.
The values $s_1$, $s_2$, $\ldots$ are distributed as in $(A.4)$.
The values $y_1$, $y_2$, $\ldots$ are distributed as follows.
Given that the path has jumped from the origin of branch
$y_k$, the distribution of the value $y_{k+1}$ is
$$\hbox{Prob}\left(\nu<y_{k+1}<\nu+d\nu\right) =
f (y_{k}, \nu)\,d\nu\ \ .\eqno(A.13)$$
The paths $\bar q$ are defined as in $(A.7)$ with $k_j$ replace by
$y_j$.  The functional integral for this system is
$${\cal A}^{\rm B}_{y_i,y_f}\left(x_i, x_f,\tau\right) = \int
[d\bar q]_{\rho,f}\,e^{- {1\over 2}\int^\tau_0 \dot{\bar q}^2 d\tau'}\,
e^{-\sum\limits^{\#}_{k=0} \alpha(y_{k}) \ell_k}
e^{i\sum\limits^{\#-1}_{k=0}h\left( y_{k}, y_{k+1}\right)}\eqno(A.14)$$
where the path integral is over all paths $\bar q$ from $x_i$
on branch $y_i$ to $x_f$ on branch $y_f$ in time $\tau$.
Note that $y_0\equiv y_i$, $y_\#\equiv y_f$ and the function $f$, $\rho$,
$h$ and $\alpha$ are the same functions introduced
in section~II.
Equivalently the path integral can be perform by first integrating over
all values $s_1$, $y_1$, $s_2$,  $y_2$, $\ldots$ restricted to
$$\sum\limits^{\#}_{k=1} (y_k-y_{k-1})=y_f-y_i\eqno(A.15)$$
with $q^*_x$ fixed and then integrating over all paths $q^*_x$.
We can see now that the path integral
${\cal A}_{y_i,y_f}$ in (3.31)
is given by the path integral (A.14).  We can show that
the functional integral $(A.14)$ satisfies
$${d\over dx_f}{\cal A}^{\rm B}_{y_i,y_f}(x_i,x_f)\bigg|_{x_f=0}
= \int^{+\infty}_{-\infty} dy_f' g(y_f',y_f){\cal A}^{\rm
B}_{y_i,y_f'}(x_i,0)\eqno(A.16)$$
which is the generalization of $(A.11)$.
\goodbreak
\bigskip
\vfill
\eject
\noindent{\bf Figure Captions}
\medskip
\noindent
Fig. 1: Quantum system made out of N-half-lines joined together
at the origin.  As an example of a particle trajectory, a particle can start
at $x_i$  on branch $1$ and can end up at $x_f$ on branch $3$.
\vfill
\end